\documentclass{article}
\usepackage[utf8]{inputenc}
\usepackage{algorithm}
\usepackage{algorithmic}
\usepackage{natbib}
\let\cite\citep
\usepackage{amsmath}
\usepackage{amsfonts} 
\usepackage{amssymb}
\usepackage{amsthm}
\usepackage{romannum}
\usepackage{graphicx}

\usepackage{color}
\usepackage{soul}

\usepackage{fullpage}
\newcommand{\single}{{single}}
\newcommand{\mult}{{multi}}
\newcommand{\multi}{\mult}

\newcommand{\calV}{\mathcal{V}}
\newcommand{\calN}{\mathcal{N}}
\newcommand{\E}{\mathbb{E}}
\newtheorem{example}{Example}
\newtheorem{theorem}{Theorem}[section]
\newtheorem{definition}[theorem]{Definition}

\newcommand{\argmax}{\mathop{\rm argmax}\limits}

\title{Who Benefits from a Multi-Cloud Market? A Trading Networks Based Analysis}
\author{
  Segev Wasserkrug\footnote{IBM Research - Israel.  {\tt segevw@il.ibm.com}}
  \ \
  Takayuki Osogami\footnote{IBM Research - Tokyo.  {\tt osogami@jp.ibm.com}}
  \ \
}
\date{}

\begin{document}
\pagenumbering{arabic}

\maketitle

\begin{abstract}
    In enterprise cloud computing, there is a big and increasing investment to move to multi-cloud computing, which allows enterprises to seamlessly utilize IT resources from multiple cloud providers, so as to take advantage of different  cloud providers' capabilities and costs. This investment raises several key questions: Will multi-cloud always be more beneficial to the cloud users? How will this impact the cloud providers?  Is it possible to create a multi-cloud market that is beneficial to all participants?

In this work, we begin addressing these questions by using the game theoretic model of trading networks and formally compare between the single and multi-cloud markets. This comparison a) provides a sufficient condition under which the multi-cloud network can be considered more efficient than the single cloud one in the sense that  a centralized coordinator having full information can impose an outcome that is strongly Pareto-dominant for all players and b) shows a surprising result that without centralized coordination, settings are possible in which even the cloud buyers' utilities may decrease when moving from a single cloud to a multi-cloud network. As these two results emphasize the need for centralized coordination to ensure a Pareto-dominant outcome and as the aforementioned Pareto-dominant result requires truthful revelation of participant's private information, we provide an automated mechanism design (AMD) approach, which, in the Bayesian setting, finds mechanisms which result in expectation in such Pareto-dominant outcomes, and in which truthful revelation of the parties' private information is the dominant strategy. We also provide empirical analysis to show the validity of our AMD approach.
\end{abstract}

\section{Introduction}
In the enterprise cloud computing market, there is an increasing trend to move to \textit{multi-cloud} computing, in which enterprises can seamlessly distribute their computational tasks among multiple cloud providers, so as to take advantage of the different capabilities and costs of these providers. Such a multi-cloud market intuitively seems to benefit the cloud users while being detrimental to existing large cloud providers. As most of today's cloud resources are provided by such large providers, it will be difficult to realize the multi-cloud vision without their support. Therefore, in order to understand, from an economic standpoint, the benefits and pitfalls of the multi-cloud market, there is a need to formally analyze, exactly who, and under which conditions, will benefit or lose from moving to a multi-cloud market. In addition we would like to understand whether a multi-cloud market is possible that is beneficial to all participants. 

To carry out this analysis requires formally modeling and comparing between the existing single cloud market and desired multi-cloud one. As cloud markets are two sided markets between the cloud users (or buyers) and the cloud providers, between which there are contracts, a model is required which captures the need for the buyers and providers to agree on the prices paid for the IT infrastructure. Therefore, rather then modeling this as a normal form game, which has been the approach taken in many previous applications of game theory to cloud computing, we model these markets as \emph{trading networks} \cite{trading13}, which model the outcome of a game as a set of contracts, and in which the  accepted solution concept is a \textit{stable outcome},  which requires joint agreement between the parties on the set of preferred contracts (a formal definition of stable outcomes appear in the sequel). To the best of our knowledge, this is the first usage of trading networks in the context of cloud computing

Using the trading networks model, we carry out a comparative analysis of the two markets. Our first result formalizes the intuition that a multi-cloud market is more efficient than a single cloud one, by showing that there is a formal sufficient condition such that, whenever this condition holds, a centralized coordinator that has access to the parties' private information can create an outcome that is strongly Pareto-dominant for all cloud users and providers.  However, our analysis also shows the more surprising result that in the absence of such centralized coordination, not only are such Pareto-Dominant outcomes not guaranteed, but that there are cases in which even the cloud \textbf{buyers} utilities' may decrease when moving to a multi-cloud setting. 

The fact that cloud users' utilities may decrease when moving to the multi-cloud market strengthens the need for a centralized market coordinator. However, the constructive proof used to create the Pareto-dominant multi-cloud outcome requires both cloud users and cloud providers to reveal their information technology (IT) infrastructure costs, which are typically proprietary information. Moreover (see  \citet{AMDTradingijcai2023}), in trading network settings, parties are often incentivized to be untruthful about such costs, and even when adding a centralized coordinator, such that all contracts and payments go though that coordinator, it is impossible to design a priori even for simple trading networks a mechanism which is ex-post a) efficient, b) one in which the dominant strategy for all participants is to be truthful (formally called \textit{dominant strategy incentive compatible} or \textit{DSIC}), c) the coordinator's utility from participating in the network is not negative (formally, \textit{weakly budget balanced} or \textit{WBB}) and d) that the utility of all parties is non-negative (formally, \textit{individually rational} or \textit{IR} - which is a special case of Pareto-dominance). Therefore, we build upon \textit{Automated Mechanism Design} (\textit{AMD}) work in \citet{AMDTradingijcai2023} to show how mechanisms can be created, for some distributions of cost functions, which are ex-post DSIC and efficient, ex-ante WBB, and Pareto-dominant in expectation. 

To summarize our contributions: we model both the single cloud and multi cloud market using the trading network model, formally show that  the multi-cloud market is more efficient given a centralized coordinator and that without such coordination, there are settings in which the cloud users' utilities may decrease when moving to a multi-cloud environment. Furthermore, we provide an AMD approach which, for many IT cost distributions, results, in expectation, in a strongly Pareto-dominant market and in which truthful participation for all parties is the dominant strategy. Finally, we provide empirical analysis which demonstrates the validity of our AMD approach.

\section{Related Work}
The potential benefits of moving to the multi-cloud market, as well as the technical capabilities required for this move, appear in many works \cite{sky09,supercloud15}. \citet{SupercloudEconomics2016} discuss at a high level the economics of multi-cloud for both buyers and providers. \citet{https://doi.org/10.48550/arxiv.2205.07147} provide a contemporary and comprehensive view on multi-cloud, called \emph{Sky Computing}, and suggest implementing software components called \emph{brokers} for the distribution across multiple clouds. This work also recognizes the possible opposition of  cloud providers. However, none of the above works carry out a formal market analysis or design.

A large body of previous works exists on the application of game theory to cloud computing. This includes: pricing schemes for selling cloud services \cite{FixedandMarketPricing2012,SimpleCloudPricingSchemes2019}, truthful online scheduling \cite{TruthfulOnlineScheduling2022}, analysis of ad services and resulting network effects \cite{AdCloudServicesNetworkEffects2018}, spot pricing of cloud instances \cite{SpotPricingCloud2016}, a game theoretic analysis of the benefits of cloud computing \cite{CloudComputingBenefitsGameTheory2012}  and a mean field analysis of cloud resource sharing \cite{MeanFieldCloudResourceSharing2016}. However, none of these works have a formal analysis and comparison of the single and multi-cloud market, nor suggest any approach of moving from one market to another while increasing the utilities of all participants. Moreover, none of these works use trading networks to model the cloud settings.

Trading networks, the formal framework we use to analyze the two cloud markets, along with associated models, have been covered extensively in  a large body of work  \cite{ostrovsky2008stability,hatfield2011stability,hatfield2012matching,trading13,hatfield2015chain}. 

AMD \cite{AMDSandholm2002,AMDSandholm2003} is a class of optimization and machine learning based approaches to create use case specific mechanisms in cases in which manual mechanism design face challenges. AMD approaches generate mechanisms for specific players' types distributions,  that  satisfy desired properties such as DSIC, efficiency, IR and WBB. Most previous works on AMD such as \citet{rahme2021auction,michaelj2022auctionlearningas,pmlr-v97-duetting19a,MJG18} focus on combinatorial auctions. A work which does address AMD in trading networks is \citet{AMDTradingijcai2023}. We build upon this work to define an AMD approach to create a Pareto-dominant multi-cloud market. 

\section{Trading Networks}
In this section, we briefly summarize the trading network model as presented in \citet{trading13} and the stable outcome solution concept.

A trading network is  a tuple $TR=(\calN,\Omega,v)$, where $\calN$ is a set of players
(or firms), $\Omega$ is a set of bi-lateral \textit{trades} such that each $\omega \in \Omega$ has a buyer $b(\omega)$ and a seller $s(\omega)$, and 
$v\equiv (v_i)_{i\in\calN}$, where each $v_i:2^\Omega \rightarrow \mathbb{R}$ is the value player $i$ associates with any set of trades. The set of trades $\Omega$ defines a set of possible \textit{contracts} $X$, where $x \in X$ is of the form $x=(\omega,p_{\omega})$, where $\omega \in \Omega$ and $p_\omega \in \mathbb{R}_{\geq 0}$ is the price paid by $s(\omega)$ to $b(\omega)$ for the trade $\omega$. An outcome in a trading network is a set of \textit{feasible} contracts $Y \subseteq X$ where $Y$ is feasible if it does not contain two contracts in which the same trade appears. The utility of  $i \in \calN$ for a set of contracts $Y$ is defined by 
\begin{equation}
\label{eq:trading-network-utility}
u_i(Y) = v_i(\tau(Y))+\sum_{(\omega,p_\omega) \in s_i(Y)}p_\omega -\sum_{(\omega ,p_{\omega}) \in b_i(Y)}p_\omega
\end{equation}
where $\tau(Y) \equiv \{\omega:(\omega,p_\omega) \in Y\}$  is the set of trades in $Y$, $s_i(Y)$ is the subset of contracts in $Y$ in which $i$ is a seller and $b_i(Y)$ is the subset of contracts in $Y$ in which $i$ is a buyer.

The standard solution concept in trading networks is the \emph{stable outcome}. Similar to Nash Equlibria, a  stable outcome is a state from which players do not wish to deviate. Stable outcomes are defined based on the notion of \emph{choice correspondence} \cite{trading13} which, for each agent $i$ and set of contracts $Y \subseteq X$, are the subsets of contracts in $Y$, that maximize $i$'s utility:
\begin{align*}
C_i(Y) \equiv \argmax _{Z \subseteq Y; Z \mbox{:feasible}}u_i(Z)
\end{align*}
A stable outcome  is then defined as:

\begin{definition}[Stability (from \citet{trading13})]
\label{def:stable-outcome}
An outcome $A \subseteq X$ is \textit{stable} if it is:
\begin{enumerate}
\item \textit{Individually rational}: $A_i \in C_i(A)$ for all $i$;
\item \textit{Unblocked}: There does not exist a nonempty \textit{blocking set} $Z \subseteq X$ such that
\begin{enumerate}
\item $Z \cap A = \emptyset$, and
\item for every $i \in b(Z) \cup s(Z)$ for every $Y \in C_i(Z \cup A)$, we have $Z_i \subseteq Y$, where $Z_i$ is the set of trades in $Z$ that involve $i$ as the buyer or seller.
\end{enumerate}
\end{enumerate}
\end{definition}

Intuitively, this means that when presented with a stable outcome $A$, one cannot propose a new set of contracts such that all the agents involved in the new contracts would \textbf{strictly} prefer forming all of them (and possible dropping some of existing contracts in $A$) to sticking with $A$. 

\section{\label{sec:modeling}Modeling Cloud Computing}
\subsection{The Cloud Market}

A cloud market has a set of \emph{cloud buyers} which are the potential users of the cloud resources, and which we denote by $CB$, such that each cloud buyer $cb \in CB$ has a set of \emph{computational workflows} they need to run, where  each workflow of buyer $cb$ is composed of a set of \emph{tasks} $T_{cb}$. An example of such a workflow is a machine learning workflow, which could be composed of a data extraction task and a machine learning task that uses deep neural networks. Each such task requires a set of resources.  In this machine learning workflow, the data extraction task requires access to the database where the data resides, and the deep neural network machine learning task will require access to GPUs. 

In a cloud market, there is also a set of \emph{cloud providers} $CP$. Each provider $cp \in CP$ has a set of computational resources such as storage, GPUs and software services, which can be provided to buyers to fulfill their tasks. In our model,  for any buyer $cb\in CB$, we assume that each of the tasks of $cb$ can be implemented either on computational resources provided by one of the providers $cp \in CP$, or on internal computational infrastructure owned and operated by $cb$. Buyers pay the providers for providers' computational resources they use, and have costs associated with the internal computational resources. Formally, we will model the cost of a buyer $cb$ for the internal computational resources for a subset of tasks $T' \subseteq T_{cb}$ by a function  $F_{cb}:2^{T_{cb}} \rightarrow \mathbb{R}^+$, and the cost of provider $cp \in CP$ for providing set of computational resources for a subset of tasks $T' \subseteq T \equiv \cup_{cb \in CB} T_{cb}$ by a function $c_{cp}:2^T \rightarrow \mathbb{R}^+$ (these different functions encapsulate the cost and capabilities differences between providers). For simplicity of exposition, we assume that $\forall cp \in CP$, $\forall cb \in CB$,  $c_{cp}(\emptyset)=F_{cb}(\emptyset)=0$, i.e., the cost of not providing any resources for the providers or using any internal resources for the buyers is zero\footnote{In practice, cloud providers have infrastructure in place which results in costs, even if this infrastructure is not used. It is easy to incorporate such costs, for example, by ensuring that $v_{cp}(\emptyset)$ is negative for $cp \in CP$. However, for the remainder of this work, we ignore these costs and assume that $\forall cp \in CP$, $v_{cp}(\emptyset)=0$.}.  Finally, we assume that for each $cb \in CB$, there is a value $V_{cb}$ for $cb$ for running the tasks $T_{cb}$, such that $V_{cb} \geq F_{cb}(T_{cb})$.

To model both a single cloud and multi-cloud market, we define a trading network  $TR=\{ \calN_c, \Omega,v\}$  where:
\begin{itemize}
\item The agents are the set $\calN_c=CB \cup CB$  \footnote{While not detailed in this work, the trading networks model and our major results also cover more complex cloud markets such as ones including  organizations that provide higher level cloud services to buyers on top of other cloud providers' infrastructure (\citet{CloudEconomics2017} and \citet{CSVP2016}), i.e., organizations which are both buyers and sellers.}.
\item The set of possible trades is  $\Omega \equiv \{(cp,cb,T'_{cb}):  cp \in CP, cb \in CB, T'_{cb} \subseteq T_{cb}\}$, i.e., there is a trade for each combination of provider, buyer, and subset of tasks for that specific buyer. 
\item $v = v_{CP} \cup v_{CB}$ where $\forall \Psi \subseteq \Omega$, $v_{cp} \in v_{CP}$ is such that $v_{cp}(\Psi) = -c_{cp}(T_{cp}^{\Psi})$ where $T_{cp}^\Psi=\cup_{(cp,cb,T'_{cb})\in \Psi}T'_{cb}$ is the set of tasks in $\Psi$ for which $cp$ is a seller. I.e.,  $v_{cp}(\Psi)$ is $cp$'s cost for providing the resources required by all tasks appearing in the trades $\Psi$. Note that we assume that if $cp$ cannot provide the resources for  $ T_{cp}^\Psi$,  then $c_{cp}(T_{cp}^\Psi)=\infty$.
\end{itemize}

The above definitions are common for both single cloud and multi-cloud networks. The differences between the two settings is captured by the valuation function $v$ of the buyers as follows

\begin{itemize}
\item In the single provider setting, the value of selecting resources from more than one provider is $-\infty$. Formally, if $\exists cp_1,cp_2 \in CP$ and $\exists cb \in CB$ such that for some task $t \in T_{cb}$ both $t \in T_{cp_1}^\Psi$ and $t \in T_{cp_2}^\Psi$  (we call this \emph{Condition 1}), then $v_{cb}(\Psi)=-\infty$.

\item In the multi-cloud setting, or in the single provider setting where Condition 1 does not hold, $v_{cb}(\Psi)=V_{cb}-F_{cb}(T_{cb} \setminus T_{cb}^\Psi) $ where $T_{cb}^\Psi=\cup_{(cp,cb,T'_{cb})\in \Psi}T'_{cb}$ .  Note that $F_{cb}(T_{cb} \setminus T_{cb}^\Psi) $ is  the cost for buyer $cb$ of implementing internally the tasks that are \emph{not} fulfilled by trades in $\Psi$.
\end{itemize}

Given the above, and  an outcome defined by a set of contracts $Y$, we define the utility of each agent from $Y$ according to Equation \eqref{eq:trading-network-utility}, i.e.:
\begin{itemize}
\item For each buyer $cb$, $u_{cb}(Y)=v_{cb}(\tau(Y)) - \sum _{(\omega,p_{\omega})\in Y \text{ s. t. } b(\omega)=cb} p_{\omega}$
\item For each provider $cp$,  $u_{cp}(Y)=v_{cp}(\tau(Y)) + \sum _{(\omega,p_{\omega})\in Y \text{ s. t. } s(\omega)=cp} p_{\omega}$. 
\end{itemize}

\textbf{Comment:} The above model assumes that a provider can provide the same infrastructure to different buyers at different prices. While this may not hold in the consumer market, it does apply in the enterprise market, where confidential contracts can be agreed between providers and buyers, based on, for example, volume discounts.

Example \ref{exa:cloud-network}. describes a concrete example of  a cloud trading network.

\begin{example}
\label{exa:cloud-network}
There are two cloud buyers, $CB=\{cb_1,cb_2\}$ and two cloud providers $CP=\{cp_1,cp_2\}$. The buyers have the following tasks: $T_{cb_1}=\{t_1^1,t_2^1\}$, $T_{cb_2}=\{t_1^2\}$. $cb_1$'s costs and values are: $V_{cb_1}=130$, $F_{cb_1}(t_1^1)=F_{cb_1}(t_2^1)=60$ and $F_{cb_1}(\{t_1^1,t_2^1\})=120$. $cb_2$'s costs and values are: $V_{cb_2}=70$ and  $F_{cb_2}(\{t_1^2\})=60$. $cp_1$'s costs are: $c_{cp_1}(\{t_1^1\})=c_{cp_1}(\{t_1^2\})=c_{cp_1}(\{t_2^1\})=c_{cp_1}(\{t_1^1,t_2^1\})=c_{cp_1}(\{t_1^2,t_2^1\})=55$ and $c_{cp_1}(\{t_1^1,t_2^1,t_1^2\})=\infty$, i.e., $cp_1$ can provide resources for $t_2^1$ and either $t_1^1$ or $t_1^2$   at a cost of 55, but not simultaneously for $t_1^1$ and $t_1^2$.  $cp_2$'s cost is $c_{cp_2}(\{t_1^1\})=c_{cp_2}(\{t_1^2\})=50$  and is $\infty$ for any other nonempty subset of tasks, i.e., $cp_2$ can provide resources for either task $t_1^1$ or $t_2^1$ but not both or any other tasks.

The set of trades are $\{(cp,cb_1,T):cp \in \{cp_1,cp_2\}, T \subseteq \{t_1^1,t_2^1\}\} \cup \{(cp,cb_2,t_1^2):cp \in \{cp_1,cp_2\}\}$. A feasible contract set is the set consisting of a single contract $Y=\{((cp_1,cb_1,\{t_1^1,t_2^1\}),60)\}$, i.e., that $cb_1$ uses $cp_1$'s cloud resources for both its tasks for a price of 60, and $cb_2$ only uses internal computational resources. For this set of contracts, $u_{cb_1}(Y)=V_{cb_1}-F_{cb_1}(\emptyset)-\sum_{(\omega,p_\omega) \in Y \mbox{ s.t. } b(\omega)=cb_1}p_\omega=130-0-60=60$ and $u_{cp_1}(Y)=c_{cp_1}(\{t_1^1,t_2^1\}+\sum_{(\omega,p_\omega) \in Y \mbox{ s.t. } s(\omega)=cp_1}=-55+60=5$. As there are no contracts in $Y$ in which $cb_2$ or $cp_2$ participate, we have that $u_{cp_2}(Y)=0$ and $u_{cb_2}(Y)=70-F_{cb_2}(\{t_1^2\})=10$.  
\end{example}

In the remainder of this work, we will use the following  notation: given a trading network   $TR^{\single}=(\calN_c, \Omega,v^{\single})$ in the single cloud setting, we will denote the corresponding network in the multi-cloud setting by   $TR^{\multi}=(\calN_c, \Omega,v^{\multi})$, i.e. the trading network with the same buyers, providers and trades (which encapsulate the same set of tasks), with the difference in the value functions $v_{CB}$ of the buyers between the two networks.

\section{Comparing the two cloud markets}
\label{sec:efficiency-comparison}

Modeling the respective cloud markets as trading networks allows us to formalize the intuition that the multi-cloud market is more efficient than the single cloud market. From the above model it is clear that, for every set of feasible contracts $Y^{\single}$ in the single cloud trading network $TR^{\single}=(\calN_c, \Omega,v^{\single})$, there is a feasible set of contracts $Y^{\multi}$ in the corresponding multi-cloud  trading network $TR^{\multi}=(\calN_c, \Omega,v^{\multi})$ such that:
\begin{equation}
\forall i \in \calN_c, \mbox{ } u_i^\multi(Y^\multi)\geq  u_i^\single(Y^\single)
\end{equation}Indeed, this holds trivially for $Y^\multi \equiv Y^\single$ as, according to the respective definitions of $TR^\multi$ and $TR^\single$:
\begin{itemize}
    \item Every feasible set of contracts in $TR^\single$ is feasible in $TR^\multi$, i.e. $X^\multi \supseteq X^\single$ (where $X^\multi$ is the set of all possible contracts in $TR^\multi$, and $X^\single$ is the set of all possible contracts in $TR^\single$).
    \item By definition of the value functions $v_i$, $i \in \calN_c$, $\forall \Psi \subseteq \Omega$ (where $\Omega$ are the set of trades), $v_i^\multi(\Psi) \geq v_i^\single(\Psi)$ and hence for every set of contracts $Y^\single$,  $u_i^\multi(Y^\single) \geq u_i^\single(Y^\single)$.
\end{itemize}
  However, we can also show a stronger result as follows:

\begin{theorem}
\label{theorem:multi-cloud-efficiency}
If
\begin{equation}
\label{eq:more-efficient-mc}
    \max_{\Psi \subseteq \Omega}\sum_{i \in \calN_c}v_i^\single(\Psi) < \max_{\Psi \subseteq \Omega}\sum_{i \in \calN_c}v_i^\multi(\Psi),
\end{equation}
then for every set of feasible contracts $Y^{\single}$ in the single cloud trading network $TR^{\single}=(\calN_c, \Omega,v^{\single})$ there is a set of trades $\Psi^* \subseteq \Omega$ and a set of payments $\{c_i: i \in \calN_c\}$, $c_i \in \mathbb{R}$, $\sum_{i \in \calN_c} c_i =0$ such that
\begin{equation}
\label{eq:Pareto-dominate-contracts}
\forall i \in \calN_c, \mbox{ } v_i^\multi(\Psi^*) + c_i>  u_i^\single(Y^\single).
\end{equation}
\end{theorem}

The significance of this theorem is the following: By definition of the value functions $\{v_i\}$, for any set of trades $\Psi$ the sum $\sum_{i \in \calN_c}v_i(\Psi)$  is directly proportional to the sum of the costs of the computational infrastructure for all buyers and providers. Therefore, when Equation \eqref{eq:more-efficient-mc} holds, it means that a set of trades can be found in the multi-cloud network in which the sum of all computational resource costs for all network participants is strictly lower than the sum of all costs for any set of trades in the corresponding single cloud network. What this theorem then states is that when the multi cloud network is more efficient in the sense of Equation \eqref{eq:more-efficient-mc}, for any set of contracts $Y^\single$  in the single cloud setting, there can be found a set of trades in the corresponding multi-cloud setting with (possibly negative) payments $c_i$, such that when the total payment $c_i$ is made to $i$, $i$'s utility  Pareto dominates $Y^{\single}$, i.e., all parties can benefit from this higher efficiency of the multi cloud. In addition, as $\sum_{i \in \calN_c} c_i =0$, the payments can be viewed as total payments between the cloud providers and cloud buyers, i.e., no additional payments need to be added from an external source.  Such payments could be distributed, through, for example, a centralized coordinator, which collects all  payments from the cloud buyers and providers and distributes all positive payments.

\textbf{Proof Sketch of Theorem \ref{theorem:multi-cloud-efficiency}} (full proof appears in the technical appendix): a key observation is that  in any trading networks, the sum of the valuations over a set of trades is equal to the sum of utilities over any set of contracts over these trades, i.e. $\forall Y \subseteq X$, $\sum_{i \in \calN}v_i(\tau(Y))=\sum_{i \in \calN} u_i(Y)$. Therefore, when Equation \eqref{eq:more-efficient-mc} holds, there is a positive difference between the maximum sum of utilities in the multi-cloud setting and the sum of the utilities for any contract set $Y^\single$. Consequently, given a set of trades which maximizes the sum of valuations in the multi-cloud setting (which also maximizes the utilities),  a centralized coordinator through which all payments flow can take this positive difference in the multi-cloud setting and distribute it so that each party gets a higher utility than in the single cloud one.

Thus, with a centralized coordinator, the multi-cloud network can be more beneficial to everyone in the sense of Theorem \ref{theorem:multi-cloud-efficiency}. The question is what would be the outcome of moving to a multi cloud market without  such a centralized coordination (in which case we would expect the trading network to settle on a stable outcome as defined in Definition \ref{def:stable-outcome}).

\begin{figure}[h]
    \begin{minipage}{0.49\linewidth}
    \centering
    \includegraphics[width=\linewidth]{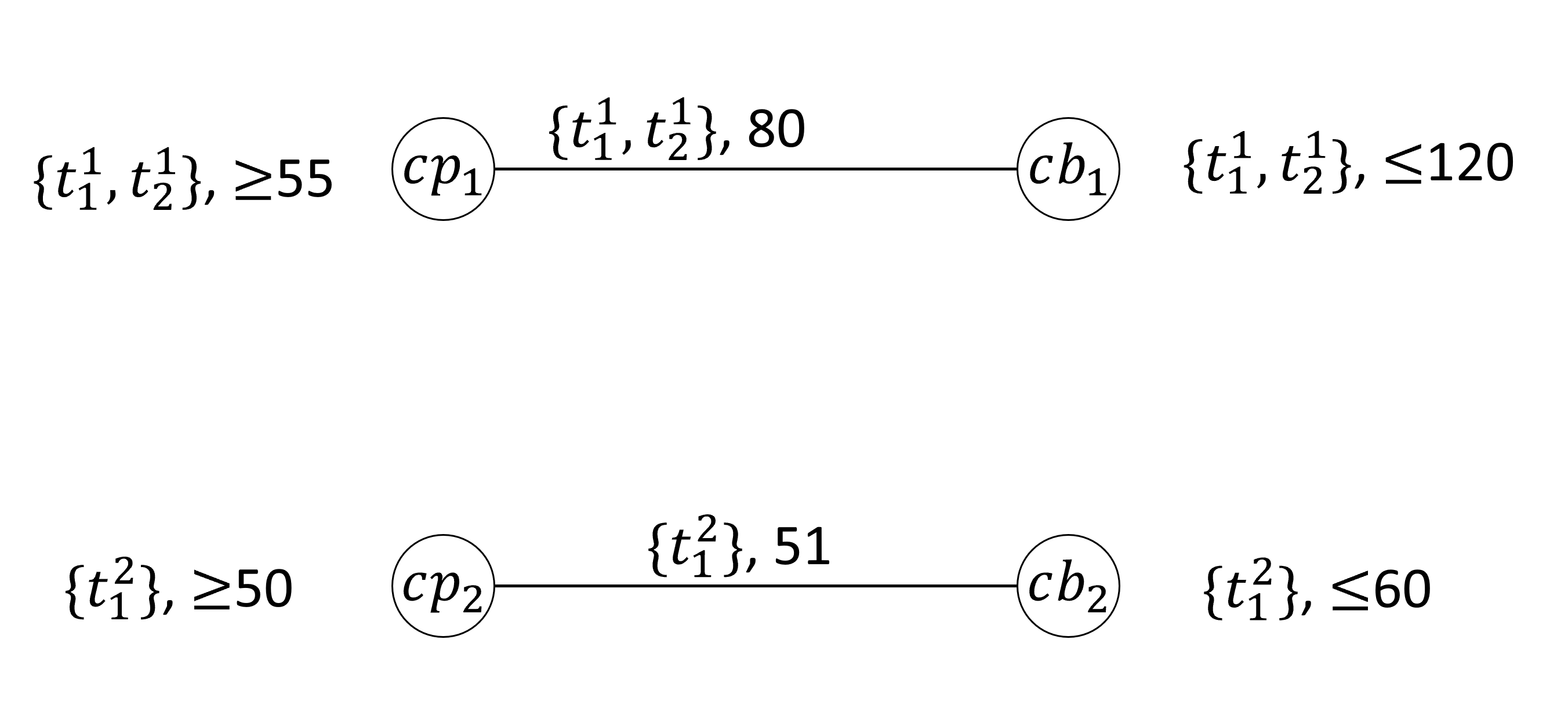}\\
    (a) Single cloud
    \end{minipage}
    \begin{minipage}{0.49\linewidth}
    \centering
    \includegraphics[width=\linewidth]{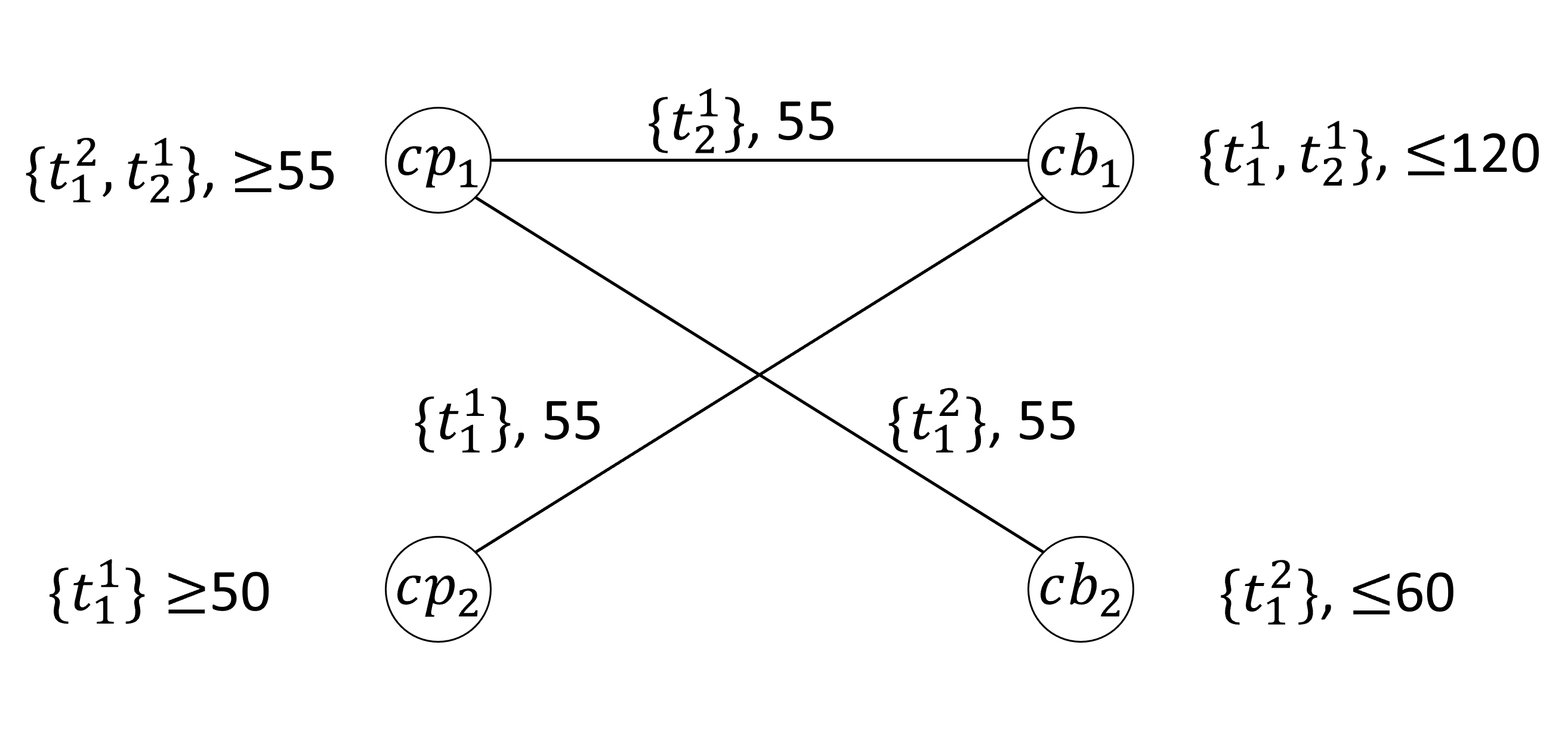}\\
    (b) Multi-cloud
    \end{minipage}
    \caption{Single (a) and multi-cloud (b) stable outcomes}
    \label{fig:single-multi}
\end{figure}
It turns out that without centralized coordination, the multi-cloud stable outcomes do not Pareto-dominate the single cloud stable outcomes \textbf{even for the buyers}. Consider Example \ref{exa:cloud-network}:   Figure \eqref{fig:single-multi} shows a stable outcome where $cb_1$ pays $80$ to $cp_1$ for $\{t_1^1,t_2^1\}$ and  $cb_2$ pays $cp_2$ 51  for $\{t_1^2\}$ in the single cloud case (Figure (\ref{fig:single-multi}a)). In Figure (\ref{fig:single-multi}b) there is a multi cloud stable outcome where $cb_1$'s tasks are split between $cp_1$ and $cp_2$ for a total cost of $110$, and $cb_2$ pays $cp_1$ $55$ for it's task, so that both buyers pay more than in the single cloud stable outcome (a proof showing that these outcomes are indeed stable appears in the technical appendix).  

\section{Pareto Dominant Multi-Cloud Mechanism}
While the multi-cloud stable outcomes do not dominate the single cloud stable outcomes, whenever Equation \eqref{eq:more-efficient-mc} holds, it is possible to create a centralized multi-cloud market which strongly Pareto-dominates the single cloud one. However, the constructive proof used to obtain such a set of payments requires full knowledge of the valuation functions $v_i$ (which, in the case of the cloud market corresponds to the $V$ values and $F$ functions of the buyers, and the $c$ functions of the providers). As these quantities include the IT costs of the cloud providers and buyers, this information is expected to be highly proprietary and confidential. Moreover, in centralized mechanism settings, trading network participants are often incentivized \textbf{not} to be truthful about their valuations (see \citet{AMDTradingijcai2023}). We therefore describe how to create a centralized mechanism  (which we will call a \textit{broker} as in \citet{sky09}) which is Pareto-dominant and in which truthful revelation of one's costs is the dominant strategy.

Formally, we would like a mechanism that, given a set of contracts $Y^\single$  in a single cloud market and a set of valuation functions $v'\equiv \{v'_i: i \in \calN_c\}$ such that $v'_i$ is the valuation function \emph{declared} by $i$, results in a set of trades $\Psi^*(v')$ and (possibly negative) payments $\{p_i(v'): i \in \calN_c\}$  from $i$ to the broker in the multi cloud market such that:
\begin{itemize}
\item 
\begin{equation}
    \label{eq:maximize-valuations}
    \Psi^* (v')\in \arg\max_{\Psi \subseteq \Omega} \sum_{i \in \calN_c}v_i'(\Psi)
\end{equation}
i.e., $\Psi^*(v')$ is a set of trades for which the sum of utilities is maximized (as maximizing the sum of $v'_i$'s maximizes the sum of utilities).
\item The mechanism is \emph{Dominant Strategy Incentive Compatible} (\emph{DSIC}),
\begin{equation}
\label{eq:DSIC}
		   \begin{split}
    \forall i \in \calN, v_i&(\Psi) +  p_i(v_i,v'_{-i}) \geq \\
                                                                   & v_i(\Psi) + p_i(v'_i,v'_{-i}), \text{ } \forall v'_i,v'_{-i}
           \end{split}
\end{equation}
where $v_i$ and $v'_i$ are the actual and reported valuation function of $i$ respectively and $v'_{-i}$ are the valuation functions reported by all players other than $i$.  This means that all players are incentivized to truthfully reveal their valuation functions irrespective of the behavior of the other players as a player never get a higher utility from revealing false valuations. 
\item \emph{$\epsilon$-Pareto-dominating} the existing set of contracts, $Y^\single$, in the single cloud setting,
\begin{equation} \label{eq:modified-ir}
    		  \begin{split}
    \forall i \in \calN, v_i&(\Psi^*(v)) +p_i (v)\geq
                         u_i^\single(Y^\single)+\epsilon
                \end{split}
\end{equation}
where $v$ are the actual valuation functions of all parties and $\epsilon\ge 0$. Namely, each participant's utility increases by at least $\epsilon$ when moving from a single cloud to a multi cloud setting. 

\item The sum of all the players' payments are nonnegative:
\begin{equation}
\label{eq:WBB}
    \sum_{i \in \calN_c}p_i (v)\geq 0.
\end{equation}
i.e., the broker does not have to add payments for Equations \eqref{eq:maximize-valuations}-\eqref{eq:modified-ir} to hold.
\end{itemize}

However, as \citet{AMDTradingijcai2023} show, it is impossible to create such a mechanism that fulfills  Equations \eqref{eq:maximize-valuations}-\eqref{eq:WBB} for all possible valuation functions $\{v_i: i \in \calN_c\}$ even for a very simple trading network with a single provider and customer, and even for the special case when $ u_i^\single(Y^\single)=0$ and $\epsilon=0$ in Equation \eqref{eq:modified-ir} \footnote{In this special case, the $\epsilon$-Pareto-dominant requirement is identical to the traditional IR requirement.}.  Therefore, following the approach in \citet{AMDTradingijcai2023}, we will address the  Bayesian setting where there is a \textit{finite set} $\calV$ o\textit{f possible sets of valuations} $v$ (also known as \textit{type spaces} for the players and a \textit{common knowledge prior distribution}\footnote{While the assumption of a common knowledge prior may seem to be overly restrictive, there are many public sources of information from which such a prior could be obtained, including the typical types of cloud workloads and their resource requirements (contributing to the buyers $V$ value), IT hardware, software and maintenance costs, and published prices by cloud providers, all contributing to the $F$  and $c$ cost functions.} $q$ over these valuations).

In this Bayesian setting, and building upon the approach in \citet{AMDTradingijcai2023}, we look for a mechanisms that finds a set of trades $\Psi\subseteq \Omega$ and a set of payments$\{p_i(v'_i,v'_{-i}): i \in \calN_c\}$ such that:
\begin{itemize}
    \item Equations \eqref{eq:maximize-valuations} and \eqref{eq:DSIC} hold, which means that the set of trades maximizes the valuations and that the mechanism is DSIC. This means that revealing the true valuation is the dominant strategy for all players, and therefore, we can assume that $\forall i \in \calN_c, \mbox{ } v'_i=v_i$.
    \item Equation \eqref{eq:WBB} is relaxed to:
\begin{equation}
\label{eq:ex-ante-WBB}
    \E[\sum_{i \in \calN}p_i(v)] \geq 0
\end{equation} where $\E$ denotes the expectation with respect to the distribution $q(\calV)$ of $v$., i.e., the broker is profitable on average over the possible valuations.
\end{itemize}

We similarly relax Equation \eqref{eq:modified-ir} to:

\begin{equation}
    \label{equ:new-IR}
    \begin{split}
    \forall {i \in \calN},  v_i \in \calV,\ &\E_{|v_i}[ v_i(\Psi^*(v)) + p_i(v)] \\
                                                                    &\geq u_i^\single(Y^\single) +\epsilon
    \end{split}
\end{equation} 
where $\E_{|v_i}$ denotes $\mathbb{E}_{q(\calV|v_i)}$ i.e., the expectation of $i$'s utility when conditioned on its true valuation $v_i$. Equation \eqref{equ:new-IR} requires that the expectation of   $i$'s utility when conditioned on its true valuation $v_i$ is at least as high as $i$'s utility in the single cloud setting\footnote{We condition on $i$'s true valuation as we assume that each participant will know what their true valuation is.}. However, Equation \eqref{equ:new-IR} raises another issue that each agent is required to reveal their true utility $u_i^\single(Y^\single)$.  As we cannot assume that revealing the true single cloud valuation is a dominant strategy,  we will assume that, rather, what is available to the broker is knowledge of the mechanism $M^\single$ by which prices are determined in the single cloud market, such as accepted bargaining processes. Formally,  we assume that $M^\single$ is a mechanism known to the broker that, given a set of valuations $v$, can calculate the utility $u_i^{M^\single}(v)$ for all $i$ in the single cloud trading trading network under $M^\single$.  Assuming this, we relax Equation \eqref{eq:modified-ir} to:
\begin{equation}
    \label{equ:new-IR-modified}
    \begin{split}
    \forall i \in \calN, v_i \in \calV,\ & \E_{|v_i}[ v_i(\Psi^*(v)) +p_i(v)]\\
                                                                 & \geq \E_{|v_i}[u_i^{M^\single}(v)] +\epsilon
    \end{split}
\end{equation}
which specifies that, conditioned on the true valuation function $v_i$, the expected utility in the multi-cloud case is higher than the expected utility in the single cloud case given mechanism $M^\single$.

To summarize, given a specific distribution $q$ over possible valuations $\calV$ (which are the providers' and buyers' cost), and a single cloud mechanism, $M^\single$, which determines the outcome  for any specific set of valuations $v\in \calV$, we wish to design a mechanism specific to $q$ and $\calV$ that maximizes the sum of valuations, is DSIC, and satisfies Equations  \eqref{eq:ex-ante-WBB} and \eqref{equ:new-IR-modified}. 

To construct such a mechanism, following again the approach outlined in \citet{AMDTradingijcai2023}, we utilize an optimization problem formulation to find a Groves mechanism \cite{Groves73} that fulfills the above requirements. A Groves mechanism is any mechanism that, given a set of reported valuations $v'$, induces a set of trades $\Psi^*(v')$ and payments $p_i(v')$ such that:
\begin{align}
   \Psi^*(v')
  \in \arg \max_{\Psi\subseteq\Omega} \sum_{i\in\calN} v'_i(\Psi) \label{eq:groves_allocation}\\
  p_i(v'_i,v'_{-i})
   = h_i(v'_{-i}) - \sum_{j\neq i} v'_j(\Psi^*(v')), \label{eq:groves_payment}
\end{align}
where $h_i$ does not depend on $i$'s valuation, $v_i$. A Groves mechanism is always ex-post efficient and DSIC. Note that a first step towards implementing a Groves mechanism requires finding an optimal set of trades $\Psi^*(v')$ that satisfies Equation \eqref{eq:groves_allocation}. We discuss how to find a set of trades in the sequel. Now, as \citet{AMDTradingijcai2023} show, given such an optimal set of trades, $\Psi^*(v')$ if $\E_{|v_i}[u_i^{M^\single}(v)] +\epsilon=0$, we can search for a set of functions $h_i$ that satisfy Equations \eqref{eq:ex-ante-WBB} and \eqref{equ:new-IR-modified} by solving the following optimization problem:

\begin{align}
    \min_h
    & \sum_{i \in \calN_c} \E\left[ h_i(v_{-i}) \right] \label{eq:objective}\\
    \mbox{s.t. }
    &
    \sum_{i \in \calN_c} \E\left[ h_i(v_{-i}) \right]
    \ge
     (|\calN_c|-1) \sum_{i \in \calN_c} \E\left[v_i(\Psi^*(v)) \right] \label{eq:WBB-Constraint}\\
    & 
     \sum_{j \in \calN_c} \E_{\mid v_i}\left[ 
        v_j(\Psi^*(v))
    \right] -  \E\left[
        h_i(v_{-i})
    \right] \ge 0, \forall i, \forall v_i,\label{eq:IR-constraint}
\end{align}
Here Constraint \eqref{eq:WBB-Constraint} enforces that the broker does not have to invest in participating, and Constraint  \eqref{eq:IR-constraint} enforces Equation \eqref{equ:new-IR-modified} when $\E_{|v_i}[u_i^{M^\single}(v)] +\epsilon=0$ .  To require that Equation \eqref{equ:new-IR-modified} holds, we can therefore modify Constraint \eqref{eq:IR-constraint}, to the following:

\begin{align}
        \E\left[
        h_i(v_{-i})
    \right] \le   
    & \sum_{j \in \calN_c} \E_{|v_i}\left[ 
        v_j(\Psi^*(v))
     -u_i^{M^\single}(v)\right] +\epsilon,\nonumber\\
     & \forall i, \forall v_i
    \label{eq:IR-constraint-modified}
\end{align}
To summarize, given as input a mechanism $M^\single$, a set of cloud buyers $CB$, a set of cloud providers $CP$ and a set of tasks $T_{cb}$ for each $cb \in CB$, the following steps are carried out to find a desired mechanism:
\begin{enumerate}
    \item A distribution over possible sets of valuation functions $\calV$ (defined by $\{F_{cb},V_{cb}: cb \in CB\}$, $\{c_{cp}:  cp \in CP\}$) e.g. based on publicly available information). 
    \item For each valuation function $v \in \calV$, an optimal set of trades, $\Psi^*(v)$  is calculated.
    \item The optimization problem defined by Equations  \eqref{eq:objective}, \eqref{eq:WBB-Constraint} and \eqref{eq:IR-constraint-modified} is solved, using appropriate values $\{\Psi^*(v): v \in \calV\}$\footnote{The specified optimization problem objective  minimizes the payments from the parties in $\calN_c$ to the broker, thereby distributing  as much value as possible between the participants. Alternatively, we can set  the objective to maximize the broker's utility.}.
\end{enumerate}
\subsection{Finding an optimal set of trades}

Implementing our AMD approach requires finding an optimal set of trades defined by Equation \eqref{eq:maximize-valuations}.  Such a set of trades can be found by solving the 0-1 integer linear programming problem defined by Equations \eqref{eq:efficient-objective}-\eqref{eq:variable-definition} \footnote{For trading networks which are \textit{fully substitutable} more efficient algorithms exist for finding an optimal set of trades. (see, e.g., \citet{CandoganFullVersionEquilibrium}). However,  a multi-cloud trading network is not necessarily fully substitutable (see technical appendix).}.

\begin{align}
    \max_{x} 
    & \sum_{i\in\calN_c} \sum_{\Psi\subseteq\Omega} c_\Psi^{(i)} \, x_\Psi \label{eq:efficient-objective}\\
    \mbox{s.t. }
    & \sum_{\Psi\subseteq\Omega} x_\Psi = 1\\
    & x_\Psi \in \{0, 1\}, \forall \Psi\subseteq\Omega \label{eq:variable-definition}
\end{align}
where $x_\Psi=1$ indicates that the set of trades $\Psi$ is selected, and $c_{\Psi}^{(i)}\equiv v_i(\Psi)$.

Note that this is an integer program with a large number of variables. Since every $cb$ can have a trade for every subset of $T_{cb}$, and each such trade can potentially have each $cp \in CP$ as a seller, we have $|\Omega|=O(|CP|\Pi_{cb \in CB}2^{|T_{cb}|})$. Therefore, the number of variables $x_\Psi$ as defined above is $2^{O(|CP|\Pi_{cb \in CB}2^{|T_{cb}|})}$.  Recall, however, that the valuation functions $v_i$ for $i \in \calN_c$ is a function only of the set of tasks assigned to each party. Therefore, all cases in which the tasks are partitioned identically between the parties result in the same value for all valuations. Due to this, we can restrict the search for an optimal set of trades over subsets of trades where each $cb \in CB$ partitions their tasks across the $|CP|$ provider and itself. For each $cb$, the number of such possible partitions is $S(|T_{cb}|,|CP|+1)$ where $S(\cdot)$ is the 2\textsuperscript{nd} Stirling number. Therefore, multiplying this across the buyers, gives us a total of $\Pi_{cb \in CB}S(|T_{cb}|,|CP|+1)$ subsets of trades (which we denote by $Part(\Omega)$), such that at least one such subset is optimal. This results in  an optimization problem with $|Part(\Omega)|=\Pi_{cb \in CB}S(|T_{cb}|,|CP|+1)$ variables, which is a much smaller number of variables:

\begin{align}
    \max_{x} 
    & \sum_{i\in\calN_c} \sum_{\Psi\in Part(\Omega)} c_\Psi^{(i)} \, x_\Psi \\
    \mbox{s.t. }
    & \sum_{\Psi \in Part(\Omega)} x_\Psi = 1\\
    & x_\Psi \in \{0, 1\}, \forall \Psi\in Part(\Omega) 
\end{align}
where $\Psi$ is one of the partitions of tasks.
\subsection{Testing the validity of the AMD approach}
To demonstrate that our approach can indeed find an $\epsilon$-Pareto-dominating multi-cloud setting in many networks,  we tried it out on many instances of networks, so as to seek, for each instance, a setting with the desired properties\footnote{While we have open sourced the code, we have not included the link here so as to maintain anonymity. We will include the link in the final version.}.   In all network instances, we consider a network where a single cloud buyer (player 0) has two tasks (task 0 and 1) to be processed, and there are three providers (player 1, 2, and 3) who can process those tasks. The network instances differed in the possible set of valuations, $\calV$, as described below.

To calculate the utility in the single cloud case, we assume that $M^\single$ results in the two tasks being covered  by the provider who has the minimum total cost (when multiple providers have the minimal total cost, we assume that the provider with the smallest index process the two tasks).  The buyer (player 0) then makes payment to the provider  who processes those tasks.  The utility of the buyer is then $-p$, where $p$ is the amount of payment.  The utility of the provider who processes the two tasks is $-c+p$, where $c$ is the total cost of processing the two tasks.  The utilities of the other providers are 0.

In the multiple provider case, we compute the mechanism that guarantees all of the desired properties by implementing and solving the linear program (LP) specified by  Equations  \eqref{eq:objective}, \eqref{eq:WBB-Constraint} and \eqref{eq:IR-constraint-modified}  (we solved the LP by using the CPLEX\textsuperscript{\textregistered} optimization engine).  

In our experiments, the possible valuation functions, $\calV$, are defined as follows: We assume that player 1 and player 2 can have one of multiple types and study two type  spaces: $\calV_1$ and $\calV_2$. Let $\calV_1 \equiv  v_1^1 \times v_2^1$, where $v_1^1$ and $v_2^1$ can take any of the four values $\{\mathrm{c_{L}c_{L}},\mathrm{c_{L}c_{H}},\mathrm{c_{H}c_{L}},\mathrm{c_{H}c_{H}}\}$ and are the respective costs of players 1 and 2.  Let  $\mathcal{V}_2 \equiv v_1^2 \times v_2^2$, where $v_1^2$ and $v_2^2$ (the respective costs of players 1 and 2) can take one of the two values $\{\mathrm{c_{L}c_{H}},\mathrm{c_{H}c_{L}}\}$.  For example, type $c_{L}c_{H}$ means that the  resource cost for task 0 is Low, and for task 1 is High; type $c_{H}c_{H}$ means that the cost is High for both tasks; other types are defined analogously.  We assume that the types of player 0 (the buyer) and player 3 (the third provider) are fixed.  Let $\mathrm{VV}$ be the type of player 0, who has Very high cost of computational resources for either task, and let $\mathrm{MM}$ be the type of player 3, who has Medium cost for either task.  Let the low cost be $c_L=2^0$ and the very high cost be $F_V=2^{10}$.  We consider all of the combinations of the medium cost $c_M$ and high cost $c_H$ such that $c_M,c_H\in\{2^1,2^2,\ldots,2^9\}$ and $c_L<c_M<c_H<F_V$.

We carried out the above experiment for a variety of random instances and studied how often we can find desirable mechanisms (indicated by having a feasible solution to the optimization problem).  We consider the amount of payment $p$ in the single provider case to be in $\{2^1,2^2,\ldots,2^{10}\}$.  For each of the 360 combinations of $(c_L,c_M,c_H,F_V,p)$, we generate 100 prior distributions over types uniformly at random, resulting in 36,000 random instances for each of the two type spaces, $\mathcal{V}_1$ and $\mathcal{V}_2$.  Table~\ref{tab:fraction} summarizes the results of the experiment.  As shown in Table \ref{tab:fraction}, for $\mathcal{V}_2$, our method has found a desirable mechanism for all of the 36,000 random instances.  For $\mathcal{V}_1$, while we found that the desired Grove mechanisms do not always exist,  our approach found desirable mechanisms for  a slim majority (52\%)  of the instances. This shows that overall, in most cases of our generated networks, a desirable mechanism can be found.

\begin{table}[th!]
    \centering
    \begin{tabular}{|r|rr|}
    \hline
        Type space & $\calV_1$ & $\calV_2$ \\
        \hline
       \% feasible instances & 52 & 100 \\
       \hline
    \end{tabular}
    \caption{\% of satisfying mechanism instances.}
    \label{tab:fraction}
\end{table}
\section{Summary and Future Work}
Using the trading networks  model, we have analyzed and compared between the single cloud and multi-cloud markets. Our model and analysis showed that the multi-cloud market is more efficient than the single cloud one, provided a characterization of when the multi-cloud market can be made strictly more efficient, and showed that without centralized coordinators, stable outcomes for the multi-cloud market do not Pareto-dominate the single cloud ones, even for the buyers.  Finally,  we have provided an AMD based approach for creating a multi-cloud market which is strongly Pareto-dominant in expectation for all participants and in which the dominant strategy is for each party to truthfully reveal their private information. We have also provided an empirical analysis, showing that for many cases, such a Pareto-dominant market can indeed be found by our method. To the best of our knowledge, ours is the first work to provide a game theoretic comparison of the two cloud markets and an approach for designing a desired multi-cloud market.  

An important direction for future work is enhancing the AMD approach to enable finding desired mechanisms for more instances, by, for example, weakening the  DSIC requirement to Bayesian Nash incentive compatibility. Another important enhancement is to increase the computational efficiency of the AMD approach, possibly by not requiring efficiency and focusing on Pareto-dominance.
\bibliographystyle{refs.bst}
\bibliography{refs}

\clearpage
\appendix

\section{Proof of Theorem 5.1}

A key observation is that in any trading network $(\calN,\Omega,V) $, for all $Y\subseteq X$ (with $X$ being the set of all possible contracts), we have that:
\begin{equation}
\label{eq:sum_utilities_equal_sum_valuations}
\sum_{i \in \calN}u_i(Y)= \sum_{i \in \calN} v_i(\tau(Y))
\end{equation}
This is as all payments between the parties cancel each other out.
Now let 
\begin{align}
    \Psi^\star
    & \in \argmax_{\Psi\subseteq\Omega} \sum_{i\in\calN_c} v_i^\multi(\Psi).
\end{align}
Then, for any feasible contracts $Y^\single$, we have
\begin{align}
    \sum_{i\in\calN_c} u_i^\single(Y^\single)
   & = \sum_{i\in\calN_c} v_i^\single(\tau(Y^\single)) \nonumber \\
   & \le \max_{\Psi\subseteq\Omega} \sum_{i\in\calN_c} v_i^\single(\Psi) \nonumber \\
  &  < \max_{\Psi\subseteq\Omega} \sum_{i\in\calN_c} v_i^\multi(\Psi) \\
   & \le \sum_{i\in\calN_c} v_i^\multi(\Psi^\star) \nonumber
    \label{eq:proof:strict}
\end{align}
where the strict inequality follows from the assumption of the theorem.

Let
\begin{align}
    c_i' & \equiv u_i^\single(Y^\single) - v_i^\multi(\Psi^\star) \\
    c & \equiv - \sum_{i\in \calN_c} c_i'\\
    c_i & \equiv c_i' + c / |\calN_c|.
\end{align}
Then
\begin{align}
    \sum_{i\in\calN_c} c_i
    & = \sum_{i\in\calN_c} c_i' + c \\
    & = 0 \\
    \intertext{and}
    c
    & = -\sum_{i\in\calN_c} c_i' \\
    & = -\sum_{i\in\calN_c} u_i^\single(Y^\single) + \sum_{i\in\calN_c} v_i^\multi(\Psi^\star) \nonumber \\
    & > 0,
    \label{eq:proof:c_positive}
\end{align}
where the inequality follows from \eqref{eq:proof:strict}.
Also, for any $i\in\calN_c$, we have
\begin{align}
    v_i^\multi(\psi^\star) + c_i \nonumber \\
    & = v_i^\multi(\psi^\star) + u_i^\single(Y^\single) \nonumber \\ 
    & - v_i^\multi(\Psi^\star) + c / |\calN_c| \nonumber \\
    & = u_i^\single(Y^\single) + c / |\calN_c| 
\end{align}
and as due to  \eqref{eq:proof:c_positive} we have that $c/|\calN_c|>0$, we have that:
\begin{align}
     v_i^\multi(\psi^\star) + c_i >u_i^\single(Y^\single)
\end{align}
.

\section{The outcomes in Figure (1) are stable}
 Figure (1a) is stable as:
 \begin{itemize}
     \item $cb_2$ will not be willing to pay more than $51$ to $cp_1$ for $t_1^2$.
     \item $cp_1$ will not be willing to accept $51$ or less, as doing so will requires $cb_1$ to pay $cp_1$ more then $30$ for $t_2^1$ so that $cp_2$ gets a higher utility and strictly prefers these contracts.
     \item However, for  $cb_1$ to have the same utility,  $cb_1$ will need to pay $cp_2$ less than $51$ for $t_1^1$, which $cp_2$ will not be willing to accept, as this reduces $cp_2$'s current utility. 
 \end{itemize}
Therefore, there are no sets of contracts strictly preferred by all parties involved.
    
    Figure (1b) is stable as  neither $cp_1$ nor $cp_2$ will agree to a set of contracts in which they earn less, and neither buyer will be willing to pay more than they currently do. 
    
\section{The Multi-Cloud Trading Network is not Fully Substitutable}

\begin{definition}[MFS from Definition 3 of \cite{trading13}]
\label{def:MFS}
The preferences of agent $i$ are \emph{matching-theory fully subsitutable} (MFS) if:
\begin{enumerate}
    \item \label{item:Condition1} for all sets of contracts $Y,Z \subseteq X_i$ such that  $Y_{i\rightarrow}=Z_{i \rightarrow}$ and $Y_{\rightarrow i} \supseteq Z_{\rightarrow i}$, and for each $Y^* \in C_i(Y)$, there exists a set $Z^* \in C_i(Z)$ such that $Y^*_{\rightarrow_i }\cap Z \subseteq Z^*$ and $Z_{i \rightarrow}^* \subseteq Y_{i \rightarrow}^*$;
    \item \label{item:Condition2} for all sets of contract $Y,Z \subseteq X_i$ such that $Y_{\rightarrow i} =Z_{\rightarrow i}$ and $Y_{i \rightarrow} \supseteq Z_{i \rightarrow}$, and for each $Y^* \in C_i(Y)$, there exists a set $Z^* \in C_i(Z)$ such that $Y^*_{i \rightarrow} \cap Z \subseteq Z^*$ and $Z^*_{\rightarrow i} \subseteq Y^*_{\rightarrow i}$ 
\end{enumerate}
\end{definition}

Now consider the network with a single cloud buyer $cb$ and a single cloud provider $cp$ such that $T_{cb}=\{t_1,t_2\}$ $F_{cb}(t_1)=6, F_{cb}(t_2)=6, F_{cb}(\{t_1,t_2\})=12$, $V_{cb}=13$m. This results in  $v_{cb}(\emptyset)=1$, $v_{cb}(\{t_1\})=v_{cb}(\{t_2\})=7$, $v_{cb}(\{t_1,t_2\})=13$.

Now consider the two contract sets:
\begin{align}
Z=\{&((cp,cb,\{t_1\}),4), \nonumber \\
    &((cp,cb,\{t_2\}),8), \nonumber \\
    & ((cp,cb,\{t_1,t_2\}),10)\} \nonumber
\end{align}
and 
\begin{align}
Y=\{&((cp,cb,\{t_1\}),4), \nonumber \\
    &((cp,cb,\{t_2\}),4),\nonumber\\
    &((cp,cb,\{t_2\}),8),\nonumber\\
    & ((cp,cb,\{t_1,t_2\}),10)\} \nonumber
\end{align}
Therefore $Y_{\rightarrow cb} \supseteq Z_{\rightarrow cb}$ and $Y_{cb \rightarrow}=Z_{cb \rightarrow}=\emptyset$.

However, $Z^*=C_{cb}(Z)=\{((cp,cb,\{t_1,t_2\}),10)\}$ and $Y^*=C_{cb}(Y)=\{((cp,cb,\{t_1\}),4),((cp,cb,\{t_2\}),4)\}$ so $Y_{\rightarrow cb}^* \cap Z = Y^*$ but $Y^* \nsubseteq Z^*$ 
, so $Y_{\rightarrow cb}^* \cap Z \nsubseteq Z^*$
, which is a contradiction of \ref{item:Condition1} above

\section{Experimental Details}

All of the experiments were run on a cloud instance with one CPU core with 4~GB memory.  For each of the 360 configurations of $(C_L,C_M,C_H,C_V,p)$, the run on 100 random instances was completed within one second, where the $i$-th random instance was generated with random seed $i-1$ for $i=1,\ldots,100$.

\begin{figure}
    \centering
    \includegraphics[width=\linewidth]{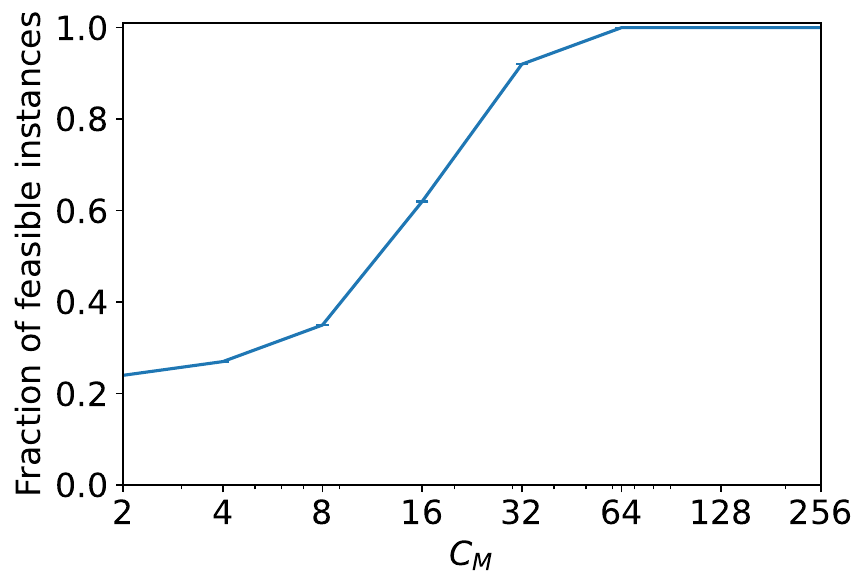}
    \caption{The fraction of feasible instances against the medium cost, $C_M$.}
    \label{fig:fraction}
\end{figure}

Table~1 in the main paper shows that 52 percent of the instances are feasible when the type space is $\calV_1$.  Here, we investigate what instances tend to be feasible.  Figure~\ref{fig:fraction} shows the fraction of feasible instances against the medium cost, $C_M$, when the type space is $\calV_1$.  Note that, for each value of $C_M$, there are multiple instances with varying values of the high cost, $C_H$, and the amount of payment, $p$.  Nevertheless, each value of $C_M$ has a single value of the fraction of feasible instances.  This means that the value of $C_M$ determines the fraction of feasible instances regardless of the values of $C_H$ and $p$ in the settings under consideration.  Overall, the instances tend to be more feasible when $C_M$ is high.

\end{document}